# The relation between tangle and local unitary invariants for pure three and four-qubit states

Xin-wei Zha[*], Yun-Guang Zhang, Jian-Xia Qi and Hai-Yang Song

School of Science, Xi'an University of Posts and Telecommunications, Xi'an, 710121, P R China

**Abstract**  A new form of local unitary (LU) transformation invariant is given for multi-qubit states . The general relation between tangle and the LU transformation invariant of pure three and four-qubit states is given. We find that the tangle actually is a special LU transformation invariant. Furthermore, the tangle is a important local unitary (LU) transformation invariant which is a nessesary condition for maximally multi-qubit entangled state.

**1. Introduction**

Entanglement is considered to be the most important resource for quantum information and computation and plays a key role in the foundations of quantum mechanics. Therefore, a great amount of theoretical effort has been performed in recent years to exploit entangled states and reveal their entanglement properties with respect to the usefulness for given quantum information tasks.

Non-locality is one of the astonishing phenomena in quantum mechanics. A common feature of entanglement for multipartite quantum systems is that the non-local properties do not change under local transformations, i.e., the unitary operations are acted independently on each of the subsystems. Hence the entanglement can be characterized by the invariants under LU transformations. Numerous researchers have investigated the equivalent classes of three-qubit states specified by LU transformation invariants [1–10], There are many other important researches to clarify the features of entanglement in multipartite systems[11–15]. In [5], A. Sudbery studied invariants of three-qubit states under LU transformations with polynomials. In [7], N. Linden *et al.* studied the invariants of multi-particle states by density matrices. It has been shown that the entanglement of three-qubit pure states is expressed by five parameters [16].

As for an arbitrary pure three-qubit state, Coffman, Kundu and Wootters have presented the three-tangle[3]. Which quantifies the genuine three-party entanglement. Recently S. Shelly Sharma et al [17] have obtained an expression for four-tangle by using simple mathematics.

---

[*] Corresponding Author. Tel: +86-29-88166094.*E-mail address:* zhxw@xupt.edu.cn (Xin-wei Zha)

In this paper, we propose a relation between local unitary transformation invariant and tangle for pure three-qubit state and four-qubit state. The tangle actually is a special LU transformation invariant.

## 2. LU equivalent and LU transformation invariants

In order to find the relation between the LU transformation invariants and the tangle, We first recall some LU transformation invariants in the following equations (1)–(4) [18,19] for multi-qubit states $|\psi\rangle$,

$$T = \langle \psi | \psi \rangle \tag{1}$$

$$F_i = \langle \psi | \hat{\sigma}_{ix} | \psi \rangle^2 + \langle \psi | \hat{\sigma}_{iy} | \psi \rangle^2 + \langle \psi | \hat{\sigma}_{iz} | \psi \rangle^2 \tag{2}$$

$$\begin{aligned} F_{ij} &= \langle \psi | \hat{\sigma}_{ix}\hat{\sigma}_{jx} | \psi \rangle^2 + \langle \psi | \hat{\sigma}_{ix}\hat{\sigma}_{jy} | \psi \rangle^2 + \langle \psi | \hat{\sigma}_{ix}\hat{\sigma}_{jz} | \psi \rangle^2 \\ &+ \langle \psi | \hat{\sigma}_{iy}\hat{\sigma}_{jx} | \psi \rangle^2 + \langle \psi | \hat{\sigma}_{iy}\hat{\sigma}_{jy} | \psi \rangle^2 + \langle \psi | \hat{\sigma}_{iy}\hat{\sigma}_{jz} | \psi \rangle^2 \\ &+ \langle \psi | \hat{\sigma}_{iz}\hat{\sigma}_{jx} | \psi \rangle^2 + \langle \psi | \hat{\sigma}_{iz}\hat{\sigma}_{jy} | \psi \rangle^2 + \langle \psi | \hat{\sigma}_{iz}\hat{\sigma}_{jz} | \psi \rangle^2 \end{aligned} \tag{3}$$

$$\begin{aligned} F_{ijk} &= \langle \psi | \hat{\sigma}_{ix}\hat{\sigma}_{jx}\hat{\sigma}_{kx} | \psi \rangle^2 + \langle \psi | \hat{\sigma}_{ix}\hat{\sigma}_{jx}\hat{\sigma}_{ky} | \psi \rangle^2 + \langle \psi | \hat{\sigma}_{ix}\hat{\sigma}_{jx}\hat{\sigma}_{kz} | \psi \rangle^2 \\ &+ \langle \psi | \hat{\sigma}_{ix}\hat{\sigma}_{jy}\hat{\sigma}_{kx} | \psi \rangle^2 + \cdots + \langle \psi | \hat{\sigma}_{iz}\hat{\sigma}_{jz}\hat{\sigma}_{kz} | \psi \rangle^2 \end{aligned} \tag{4}$$

In this paper, we introduce some 6 degree and 8 degree LU transformation invariant further.

The 6 degree LU transformation invariant is

$$S_{ij} = \det \beta_{ij} = \begin{vmatrix} \langle \psi | \sigma_{ix}\sigma_{jx} | \psi \rangle & \langle \psi | \sigma_{ix}\sigma_{jy} | \psi \rangle & \langle \psi | \sigma_{ix}\sigma_{jz} | \psi \rangle \\ \langle \psi | \sigma_{iy}\sigma_{jx} | \psi \rangle & \langle \psi | \sigma_{iy}\sigma_{jy} | \psi \rangle & \langle \psi | \sigma_{iy}\sigma_{jz} | \psi \rangle \\ \langle \psi | \sigma_{iz}\sigma_{jx} | \psi \rangle & \langle \psi | \sigma_{iz}\sigma_{jy} | \psi \rangle & \langle \psi | \sigma_{iz}\sigma_{jz} | \psi \rangle \end{vmatrix} \tag{5}$$

For simple, let us express

$$\beta_{ij}^{11} = \langle \psi | \sigma_{ix}\sigma_{jx} | \psi \rangle, \beta_{ij}^{12} = \langle \psi | \sigma_{ix}\sigma_{jy} | \psi \rangle, \beta_{ij}^{13} = \langle \psi | \sigma_{ix}\sigma_{jz} | \psi \rangle.$$

$$\beta_{ij}^{21} = \langle \psi | \sigma_{iy}\sigma_{jx} | \psi \rangle, \beta_{ij}^{22} = \langle \psi | \sigma_{iy}\sigma_{jy} | \psi \rangle, \beta_{ij}^{23} = \langle \psi | \sigma_{iy}\sigma_{jz} | \psi \rangle$$

$$\beta_{ij}^{31} = \langle \psi | \sigma_{iz}\sigma_{jx} | \psi \rangle, \beta_{ij}^{32} = \langle \psi | \sigma_{iz}\sigma_{jy} | \psi \rangle, \beta_{ij}^{33} = \langle \psi | \sigma_{iz}\sigma_{jz} | \psi \rangle$$

Then we have

$$S_{ij} = \det \beta_{ij} = \begin{vmatrix} \beta_{ij}^{11} & \beta_{ij}^{12} & \beta_{ij}^{13} \\ \beta_{ij}^{21} & \beta_{ij}^{22} & \beta_{ij}^{23} \\ \beta_{ij}^{31} & \beta_{ij}^{32} & \beta_{ij}^{33} \end{vmatrix}$$

The 8 degree LU transformation are invariants is

$$\begin{aligned}
E_{ij} &= u_i^1 u_j^1 \left( \beta_{ij}^{22} \beta_{ij}^{33} - \beta_{ij}^{23} \beta_{ij}^{32} \right) + u_i^1 u_j^2 \left( \beta_{ij}^{23} \beta_{ij}^{31} - \beta_{ij}^{21} \beta_{ij}^{33} \right) + u_i^1 u_j^3 \left( \beta_{ij}^{21} \beta_{ij}^{32} - \beta_{ij}^{22} \beta_{ij}^{31} \right) \\
&+ u_i^2 u_j^1 \left( \beta_{ij}^{13} \beta_{ij}^{32} - \beta_{ij}^{12} \beta_{ij}^{33} \right) + u_i^2 u_j^2 \left( \beta_{ij}^{11} \beta_{ij}^{33} - \beta_{ij}^{13} \beta_{ij}^{31} \right) + u_i^2 u_j^3 \left( \beta_{ij}^{11} \beta_{ij}^{32} - \beta_{ij}^{12} \beta_{ij}^{31} \right) \\
&+ u_i^3 u_j^1 \left( \beta_{ij}^{12} \beta_{ij}^{23} - \beta_{ij}^{13} \beta_{ij}^{22} \right) + u_i^3 u_j^2 \left( \beta_{ij}^{13} \beta_{ij}^{21} - \beta_{ij}^{11} \beta_{ij}^{23} \right) + u_i^3 u_j^3 \left( \beta_{ij}^{11} \beta_{ij}^{22} - \beta_{ij}^{12} \beta_{ij}^{21} \right)
\end{aligned} \quad (6)$$

where $u_k^1 = \langle \psi | \sigma_{kx} | \psi \rangle, u_k^2 = \langle \psi | \sigma_{ky} | \psi \rangle, u_k^3 = \langle \psi | \sigma_{kz} | \psi \rangle$.

### 3. The relation between tangle and local unitary invariants

3.1 For a system of three qubits, we have

$$|\varphi\rangle_{123} = a_0 |000\rangle + a_1 |001\rangle + a_2 |010\rangle + a_3 |011\rangle + a_4 |100\rangle + a_5 |101\rangle + a_6 |110\rangle + a_7 |111\rangle \quad (7)$$

and it is assumed that the wave function satisfies the normalization condition $\sum_{i=0}^{7} |a_i|^2 = 1$.

It is known [5] that the dimension of the space of orbits is 6; therefore there are six algebraically independent local invariants. Thus there is one independent LU invariant of degree 2,

$$T = \langle \psi || \psi \rangle \quad (8)$$

Because wave function satisfies the normalization condition, then we have $T = \langle \psi || \psi \rangle = 1$.

Three linearly independent quartic LU invariants are [19]

$$F_1 = \langle \psi | \hat{\sigma}_{1x} | \psi \rangle^2 + \langle \psi | \hat{\sigma}_{1y} | \psi \rangle^2 + \langle \psi | \hat{\sigma}_{1z} | \psi \rangle^2 \quad (9)$$

$$F_2 = \langle \psi | \hat{\sigma}_{2x} | \psi \rangle^2 + \langle \psi | \hat{\sigma}_{2y} | \psi \rangle^2 + \langle \psi | \hat{\sigma}_{2z} | \psi \rangle^2 \quad (10)$$

$$F_3 = \langle \psi | \hat{\sigma}_{3x} | \psi \rangle^2 + \langle \psi | \hat{\sigma}_{3y} | \psi \rangle^2 + \langle \psi | \hat{\sigma}_{3z} | \psi \rangle^2 \quad (11)$$

There is one independent LU invariant of degree 6.

$$S_{12} = \det \beta_{12} = \begin{vmatrix} \beta_{12}^{11} & \beta_{12}^{12} & \beta_{12}^{13} \\ \beta_{12}^{21} & \beta_{12}^{22} & \beta_{12}^{23} \\ \beta_{12}^{31} & \beta_{12}^{32} & \beta_{12}^{33} \end{vmatrix} \qquad (12)$$

There is one independent LU invariant of degree 8.

$$\begin{aligned} E_{12} = & u_1^1 u_2^1 \left( \beta_{12}^{22} \beta_{12}^{33} - \beta_{12}^{23} \beta_{12}^{32} \right) + u_1^1 u_2^2 \left( \beta_{12}^{23} \beta_{12}^{31} - \beta_{12}^{21} \beta_{12}^{33} \right) + u_1^1 u_2^3 \left( \beta_{12}^{21} \beta_{12}^{32} - \beta_{12}^{22} \beta_{12}^{31} \right) \\ & + u_1^2 u_2^1 \left( \beta_{12}^{13} \beta_{12}^{32} - \beta_{12}^{12} \beta_{12}^{33} \right) + u_1^2 u_2^2 \left( \beta_{12}^{11} \beta_{12}^{33} - \beta_{12}^{13} \beta_{12}^{31} \right) + u_1^2 u_2^3 \left( \beta_{12}^{11} \beta_{12}^{32} - \beta_{12}^{12} \beta_{12}^{31} \right) \\ & + u_1^3 u_2^1 \left( \beta_{12}^{12} \beta_{12}^{23} - \beta_{12}^{13} \beta_{12}^{22} \right) + u_1^3 u_2^2 \left( \beta_{12}^{13} \beta_{12}^{21} - \beta_{12}^{11} \beta_{12}^{23} \right) + u_1^3 u_2^3 \left( \beta_{12}^{11} \beta_{12}^{22} - \beta_{12}^{12} \beta_{12}^{21} \right) \end{aligned}$$

$$(13)$$

The three tangle of three qubit is defined as[20]

$$\tau_{123} = \left| \langle \psi | \sigma_{1y} \sigma_{2y} \sigma_{3x} | \psi^* \rangle^2 + \langle \psi | \sigma_{1y} \sigma_{2y} \sigma_{3z} | \psi^* \rangle^2 - \langle \psi | \sigma_{1y} \sigma_{2y} | \psi^* \rangle^2 \right|, \qquad (14)$$

By Eqs（10-15）, one can obtain the relation between tangle and LU invariants

$$\tau_{123}^2 = \left( T^2 + F_3 - F_1 - F_2 \right)^2 + 4(S_{12} \times T - E_{12}) \qquad (15)$$

That is to say, The square of three tangle is one LU invariant of degree 8.

Finally, we give the values of these invariants for some special states(all of which are taken to be normalised).

For a factorised state,

$|\varphi\rangle_{123} = a_4 |100\rangle + a_7 |111\rangle$, $F_1 = 1$, $F_2 = 1 - 4|a_4|^2 |a_7|^2$, $F_3 = 1 - 4|a_4|^2 |a_7|^2$, $S_2 = 0$, $E_4 = 0$,
$\tau_{123}^2 = \left( F_1 + F_2 - F_3 \right)^2 + S_2 - E_4 = 0$。

For a generalised GHZ state,

$|\varphi\rangle_{123} = a_0 |000\rangle + a_7 |111\rangle$, $F_1 = 1 - 4|a_0|^2 |a_7|^2$, $F_2 = 1 - 4|a_0|^2 |a_7|^2$, $F_3 = 1 - 4|a_0|^2 |a_7|^2$ $S_2 = 0$,
$E_4 = 0$, $\tau_{123}^2 = \left( 1 + F_3 - F_1 - F_2 \right)^2 + S_2 - E_4 = 16|a_0|^4 |a_7|^4$.

For the W state,

$|\varphi\rangle_{123} = a_1 |001\rangle + a_2 |010\rangle + a_4 |100\rangle$, $F_1 = 1 - 4\left( |a_1|^2 + |a_2|^2 \right) |a_4|^2$, $F_2 = 1 - 4\left( |a_1|^2 + |a_4|^2 \right) |a_2|^2$,
$F_3 = 1 - 4\left( |a_2|^2 + |a_4|^2 \right) |a_1|^2$, $S_2 = 4|a_1|^2 |a_4|^2 \left( |a_1|^2 - |a_2|^2 - |a_4|^2 \right)$,

$$E_4 = 4|a_2|^2 |a_4|^2 \left(|a_1|^2 + |a_2|^2 - |a_4|^2\right)\left(|a_1|^2 - |a_2|^2 + |a_4|^2\right),$$

$$\tau_{123}^2 = (1 + F_3 - F_1 - F_2)^2 + 4(S_2 - E_4) = 0.$$

3.2 For a system of four qubits, we have

$$\begin{aligned}|\psi\rangle_{ABCD} = & a_0|0000\rangle + a_1|0001\rangle + a_2|0010\rangle + a_3|0011\rangle \\ & + a_4|0100\rangle + a_5|0101\rangle + a_6|0110\rangle + a_7|0111\rangle \\ & + a_8|1000\rangle + a_9|1001\rangle + a_{10}|1010\rangle + a_{11}|1011\rangle \\ & + a_{12}|1100\rangle + a_{13}|1101\rangle + a_{14}|1110\rangle + a_{15}|1111\rangle\end{aligned} \quad (16)$$

The measure of four qubit entanglement of four tangle is defined as[17, 21]

$$\begin{aligned}\tau_{1234} &= |\langle\psi|\sigma_{1y}\sigma_{2y}\sigma_{3y}\sigma_{4y}|\psi^*\rangle| \\ &= |2(a_0 a_{15} - a_1 a_{14} - a_2 a_{13} + a_3 a_{12} - a_4 a_{11} + a_5 a_{10} + a_6 a_9 - a_7 a_8)|\end{aligned} \quad (17)$$

We can obtain

$$F_1 + F_2 + F_3 + F_4 + (F_{123} + F_{124} + F_{234} + F_{134}) + 8\tau_{1234}^2 = 8 \quad (18)$$

That is to say, The square of four tangle is one LU invariant of degree 4.

For a generalised GHZ state,

$$|\varphi\rangle_{1234} = a_0|0000\rangle + a_{15}|1111\rangle, \ F_1 = F_2 = F_3 = F_4 = 1 - 4|a_0|^2 |a_{15}|^2,$$

$$F_{123} = F_{124} = F_{234} = F_{134} = 1 - 4|a_0|^2 |a_{15}|^2, \tau_{1234}^2 = 4|a_0|^2 |a_{15}|^2, \text{ if } a_0 = a_{15} = \frac{1}{\sqrt{2}}, \tau_{1234}^2 = 1.$$

$$|\varphi\rangle_{1234} = \frac{1}{2\sqrt{2}}(|0000\rangle - |0011\rangle - |0101\rangle + |0110\rangle + |1001\rangle + |1010\rangle + |1100\rangle + |1111\rangle)_{1234}$$

We have $F_1 + F_2 + F_3 + F_4 = 0$, $F_{123} + F_{124} + F_{234} + F_{134} = 8$, $\tau_{1234}^2 = 0$.

**4. Conclusion**

In this paper, for three-qubit, six algebraically independent local invariants have been presented. We have shown that there exists a relation between the LU transformation invariant and the tangle for three-qubit and four-qubit. For three-qubit, three-tangle can be expressed as a function of six algebraically independent local invariants. For four-qubit, four-tangle can be expressed by invariant of degree 4. Interestingly, we find that the tangle is equle 1 for maximally three-qubit and the tangle is equle 0 for maximally four qubit entangled state. Furthermore, the approach presented can be recursively applied to obtain other invariants as well as meaningful invariants in larger systems.

Acknowledgment

This work is supported by the Shaanxi Natural Science Foundation under Contract (No. 2013JM1009) and the Program for New Century Excellent Talents in University (Grant No. NCET-12-1046).

Reference

[1] A. Acin, A.Andrianov, L.Costa ,E.Jane, J.I.Latorre and R.Tarrach, Phys. Rev. Lett. 85 (2000) 1560.

[2] A. Acin, A.Andrianov, L.Costa ,E.Jane, and R.Tarrach, J. Phys. A: Math. Gen. 34 (2001), 6725

[3] V. Coffman, J. Kundu and W.K. Wootters , Phys. Rev. A 61 (2000) 052306.

[4] M. Grassl, M. Rötteler and T. Beth, Phys. Rev. A 58 (1998) 1833.

[5] A. Sudbery, J. Phys. A: Math. Gen. 34 (2001) 643.

[6] W.K. Wootters, Rev. Lett. 80 (1998) 2245.

[7] N. Linden, S. Popescu and A. Sudbery, Phys. Rev. Lett. 83 (1999) 243.

[8] Y. Makhlin, Quant.Info.Proc. 1 (2002) 243

[9] S. Albeverio, S. M. Fei, P. Parashar and W. L.Yang, Phys. Rev. A 68 (2003) 010303.

[10] M.S. Leifer and N. Linden, Phys. Rev. A 69 (2004) 052304.

[11] Marcio F. Cornelio, Marcos C. de Oliveira, Felipe F. Fanchini, Phys. Rev. Lett. 107 (2011) 020502

[12] Thiago R. de Oliveira, Marcio F. Cornelio, Felipe F. Fanchini, Phys. Rev. A 89, (2014) 034303

[13] F. F. Fanchini, M. C. de Oliveira, L. K. Castelano, M. F. Cornelio, Phys. Rev. A 87, (2013) 032317

[14 ]H. Tajima , Annals of Physics 329 (2013) 1

[15 ] B. Kraus, Phys. Rev. Lett. 104 , (2010) 020504

[16] N.Linden, S.Popescu, Fortschr. Phys. 46 (1998) 567.

[17] S. S. Sharma and N. K. Sharma, Phys. Rev. A 79, (2009) 062323

[18] X. W. Zha, H. Y. Song and K. F. Ren, Int. J. Quantum Inf. 8 (2010) 1251

[19] X. W. Zha, C. Z. Yuan and Y. P. Zhang, Laser Phys. Lett. 10 (2013) 045201

[20] A. Osterloh and J. Siewert, Phys. Rev. A 72, (2005) 012337.

[21]A. Wong and N. Christensen, Phys. Rev. A 63, (2001) 044301

[22] Y. Yeo and W.K. Chua, Phys. Rev. Lett. 96 (2006) 060502